\documentclass[12pt]{iopart}
\usepackage{amssymb}
\usepackage[pdftex]{graphicx}
\usepackage{float}
\usepackage{overpic}
\usepackage{soul}

\providecommand{\U}[1]{\protect\rule{.1in}{.1in}}
\newcommand{\G}{\check{\cal G}}
\newcommand{\I}{\check{\cal I}}

\newcommand{\rv}{\vec{r}}

\newcommand{\sign}{\mathrm{sign}}

\newcommand{\dg}{^{\dagger}}
\newcommand{\ua}{_{\uparrow}}
\newcommand{\da}{_{\downarrow}}

\renewcommand{\epsilon}{{\varepsilon}}

\begin{document}
\title[Giant thermoelectric effects in a proximity-coupled SF-device]{Giant thermoelectric effects in a proximity-coupled superconductor-ferromagnet device}
\author{P. Machon$^1$, M. Eschrig$^2$ and W. Belzig$^1$}
\address{$^1$ Department of Physics, University of Konstanz, D-78457 Konstanz, Germany}
\address{$^2$ SEPnet and Hubbard Theory Consortium, Department of Physics, Royal Holloway, University of London, Egham, Surrey TW20 0EX, United Kingdom}
\begin{abstract}
The usually negligibly small thermoelectric effects in superconducting heterostructures can be boosted dramatically due to the simultaneous effect of spin splitting and spin filtering. Building on an idea of our earlier work [Phys. Rev. Lett. \textbf{110}, 047002 (2013)], we propose realistic mesoscopic setups to observe thermoelectric effects in superconductor heterostructures with ferromagnetic interfaces or terminals. We focus on the Seebeck effect being a direct measure of the local thermoelectric response and find that a thermopower of the order of $\sim250$ $\mu V/K$ can be achieved in a transistor-like structure, in which a third terminal allows to drain the thermal current. A measurement of the thermopower can furthermore be used to determine quantitatively the spin-dependent interface parameters that induce the spin splitting. For applications in nanoscale cooling we discuss the figure of merit for which we find values exceeding 1.5 for temperatures $\lesssim 1$K.
\end{abstract}
\submitto{\NJP}
\maketitle

Since their discovery at the beginning of the 19$^{\rm th}$ century \cite{seebeck,oersted,peltier,thomson}, thermoelectric effects have attracted continued attention in physics, as they provide the basis for a large variety of devices used in a multitude of fields in physics and engineering connected with energy management on the nano-scale. The coupling of thermal and electric transport equations is also a basic concept in non-equilibrium statistical physics, reflected e.~g. in Onsager's famous reciprocity relations \cite{onsager}. 
In the field of quantum transport, a large number of experiments have investigated the thermopower in quantum dots \cite{staring:93,moeller:98,scheibner:05}, Andreev interferometers \cite{eom:98,jiang:05}, and single atomic and molecular junctions \cite{ludolph:99, reddy:07,widawsky:12}. Thermoelectric effects in mesoscopic systems can be understood by symmetry arguments. Roughly speaking a thermoelectric response is found if the density of states $N(\epsilon)$ is asymmetric in energy, viz. $N(\epsilon)-N(-\epsilon)\ne 0$ in a range $k_BT$ ($k_B$ is the Boltzmann constant and $T$ the temperature) around the chemical potential. 

As consequence, in standard metals, described by Fermi-liquid theory, thermoelectric effects are strongly suppressed at temperatures well below the Fermi temperature, since the single-particle density of states and the scattering rates vary on a much larger energy scale than $k_B T$. We note that, in contrast, in nanoscale conductors even in the simple Landauer picture of free electrons the transmission function may depend considerably on energy, e.~g. in quantum dots or due to interaction effects, and can lead to a sizable thermoelectric effect \cite{sivan:86,beenakker:92,guttman:95}. The same holds for strongly correlated metals with strong variation of the density of states at the Fermi level, and for semiconductors where the Fermi level is near the bottom or top of an energy band.

However, with regard to thermoelectric effects in superconductors the situation is less favorable. The most widely used Bardeen-Cooper-Schrieffer (BCS) theory of superconductivity \cite{BCS} is essentially build on top of a deeply degenerate Fermi gas. This is reflected in the almost perfect electron-hole symmetry in the standard version of the theory, appropriate for conventional low-temperature superconductors, suppressing thermoelectric effects \cite{galperin:74,ginzburg:78}. On the other hand, supercurrents can interfere with thermal currents and generate a thermoelectric voltage in interferometer geometries \cite{galperin:74}. Here the effect is essentially related to the temperature dependence of the supercurrent. Later this effect was reinvestigated further in mesoscopic Andreev interferometers \cite{claughton:96,virtanen:04,titov:08} and used to explain partially the experimental findings \cite{eom:98}.

During the last decade super\-conductor-ferro\-magnet hetero\-structures have attracted a considerable interest. Driven by the prospect to create, among other perspectives, unconventional triplet pairing \cite{tokuyasu:88,bergeret:01,kadigrobov:01,halterman:01,eschrig:03}, superconducting spin-valves \cite{tagirov:99,huertas:02} or to study spin-polarized Andreev reflection 
\cite{jong:95,kashiwaya:99,zutic:00,mazin:01,lofwander:10},
many aspects of superconductor-ferromagnet heterostructures have been investigated theoretically \cite{golubov:04,eschrigkopu:04,bergeret:05,buzdin:05,zhu:10,eschrig:11,tanaka:12} and experimentally \cite{petrashov:94,lawrence:96,giroud:98,aumentado:01,ryazanov:01,kontos:02,blum:02,sellier:03,bauer:04,weides:06,robinson:06,birge:06,keizer:06,aarts:08,sprungmann:09,leksin:11,huebler:12}. 
However, the field of thermoelectricity has remained largely unexplored.

Recently, we have pointed out the possibility to generate large local and nonlocal thermoelectric effects \cite{machon:13} in heterostructures of ferromagnets and superconductors by the combined effect of induced spin splitting and spin-polarized transport. The microscopic origin can be traced back to spin-split bands due to an induced exchange field or an external magnetic field. This effect is well known \cite{fazio:99} and has first been exploited in \cite{huertas:02} to predict an absolute spin-valve in a heterostructure. 
The broken electron-hole symmetry in each spin channel leads to thermoelectric effects of equal magnitude but opposite sign for each spin direction, resulting in a vanishing effect.
The new ingredient \cite{machon:13} is to address the spin-split density of states by spin-polarized contacts, which lift the electron-hole symmetry for the transport coefficients. Hence, according to the symmetry argument, the combined action of a spin-splitting effect and a spin-polarization of the contact creates the possibility to obtain thermo\-electric effects in superconductor-ferromagnet heterostructures. We note in passing that the effect of such spin-asymmetries has been seen experimentally \cite{tedrow:71}, but the aspect of thermal currents was not discussed at that time. Further works studied the thermoelectric effect related to impurities in bulk superconductors \cite{loefwander:04, kalenkov:12} or to magnetic field or interface induced spin splitting in tunnel junctions \cite{mersevey:94,ozaeta:14}. Note that the latter case the requires comparatively large spin splittings, for which  the Clogston-Chandrasekhar limit \cite{clogston:62, chandrasekhar:62} or inhomogenous phases  \cite{fulde:64, larkin:65} might play an important role.

In the present paper we develop the idea to combine a spin-split density of states in the superconductor with spin-polarized contacts further and study the local thermoelectric effect more detailed in experimentally realizable devices of superconductors, ferromagnetic insulators, and normal metals. We consider two cases. First we study a simple two-terminal setup where the thermopower is suppressed especially at small temperatures, since energy currents trough the superconductor can only be carried by the exponentially suppressed number of quasiparticles above the gap. Second we discuss an effective two-terminal device, where the energy current is dissipated trough a second normal metal terminal and the superconductor is only used to change the spectral properties in the contact region via the proximity effect.
To quantify the thermoelectric response we study the Seebeck coefficient $S$ and find values up to $100\mu$V/K for realistic parameters.
In the context of thermoelectric effects, another interesting point to discuss is the thermoelectric figure of merit (or "$ZT$") \cite{goldsmid:54}, giving a measure of the efficiency of refrigerating devices and generators based on thermoelectric effects. As we show below we achieve values of $1.8$, which might be interesting for low temperature nano refrigeration \cite{giazotto:06}. 

\section{The Model}

\begin{figure}
\centering
\begin{minipage}[t]{0.32\linewidth}
\vspace{0pt}
\begin{overpic}[width=0.8\linewidth]{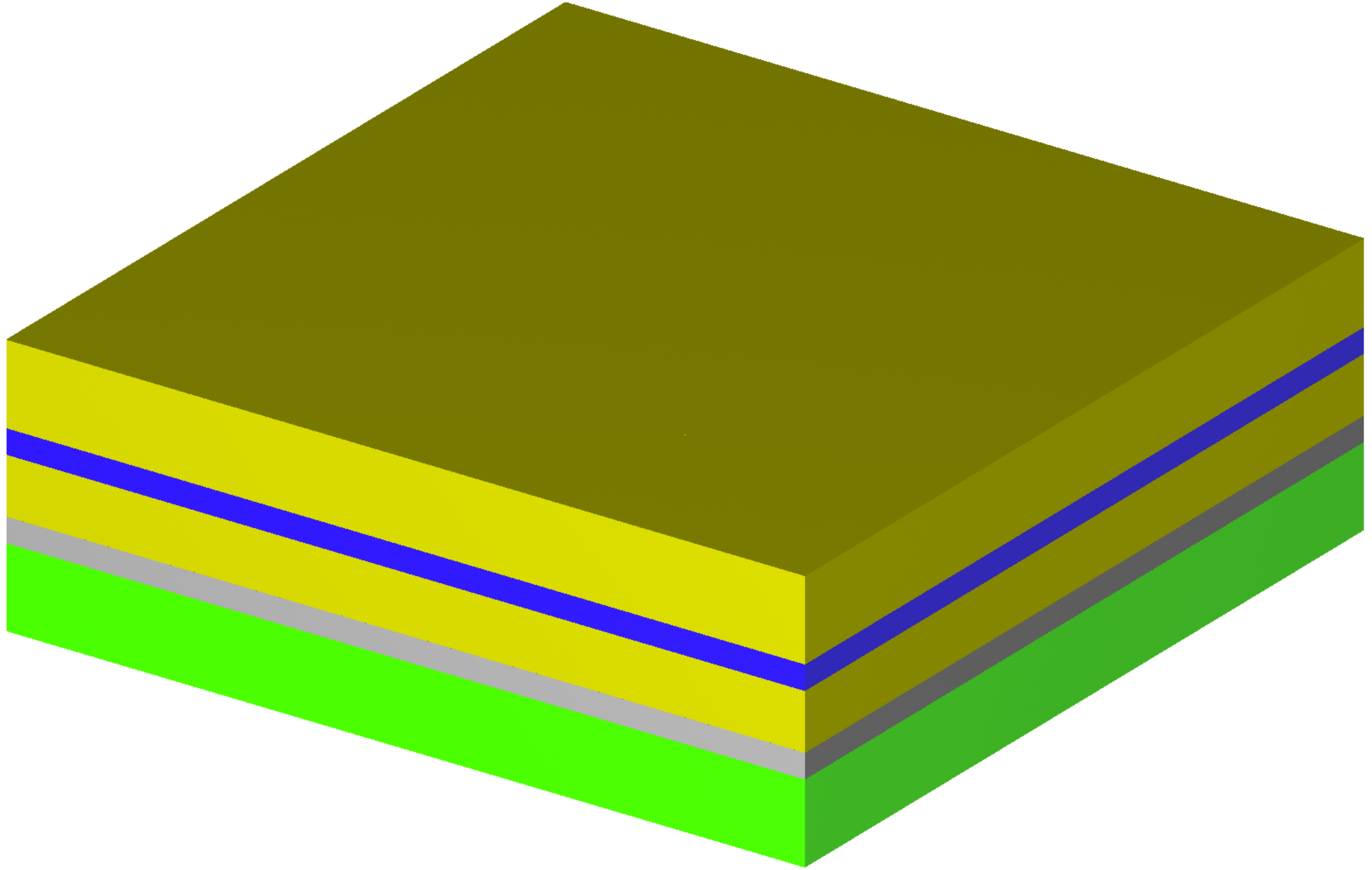}
\put(5,60){\makebox(0,3){$(a)$}}
\end{overpic}
\begin{overpic}[width=0.8\linewidth]{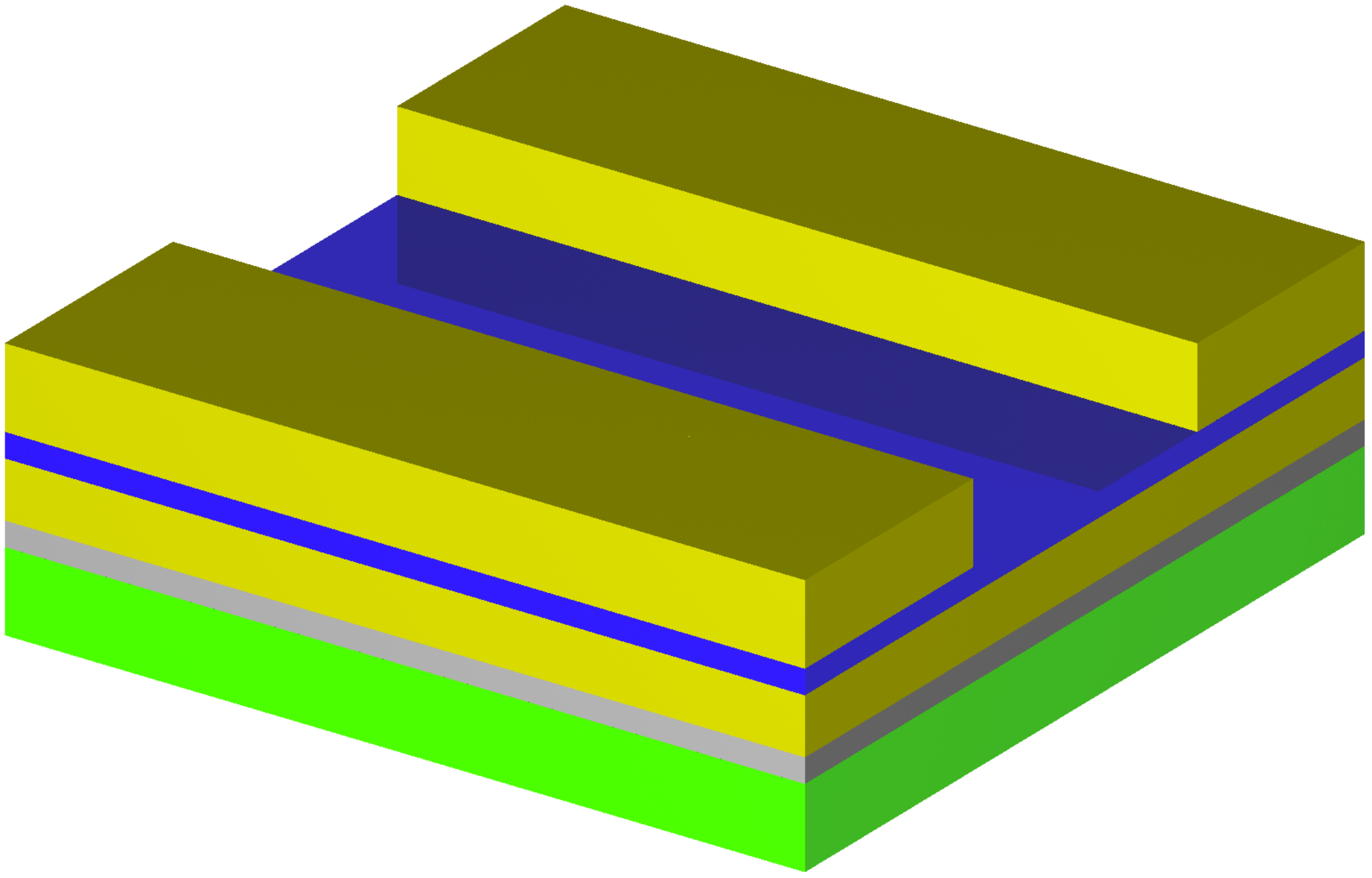}
\put(5,60){\makebox(0,3){$(b)$}}
\end{overpic}
\end{minipage} 
\begin{minipage}[t]{0.32\linewidth}
\vspace{0pt}
\begin{overpic}[width=0.99\linewidth]{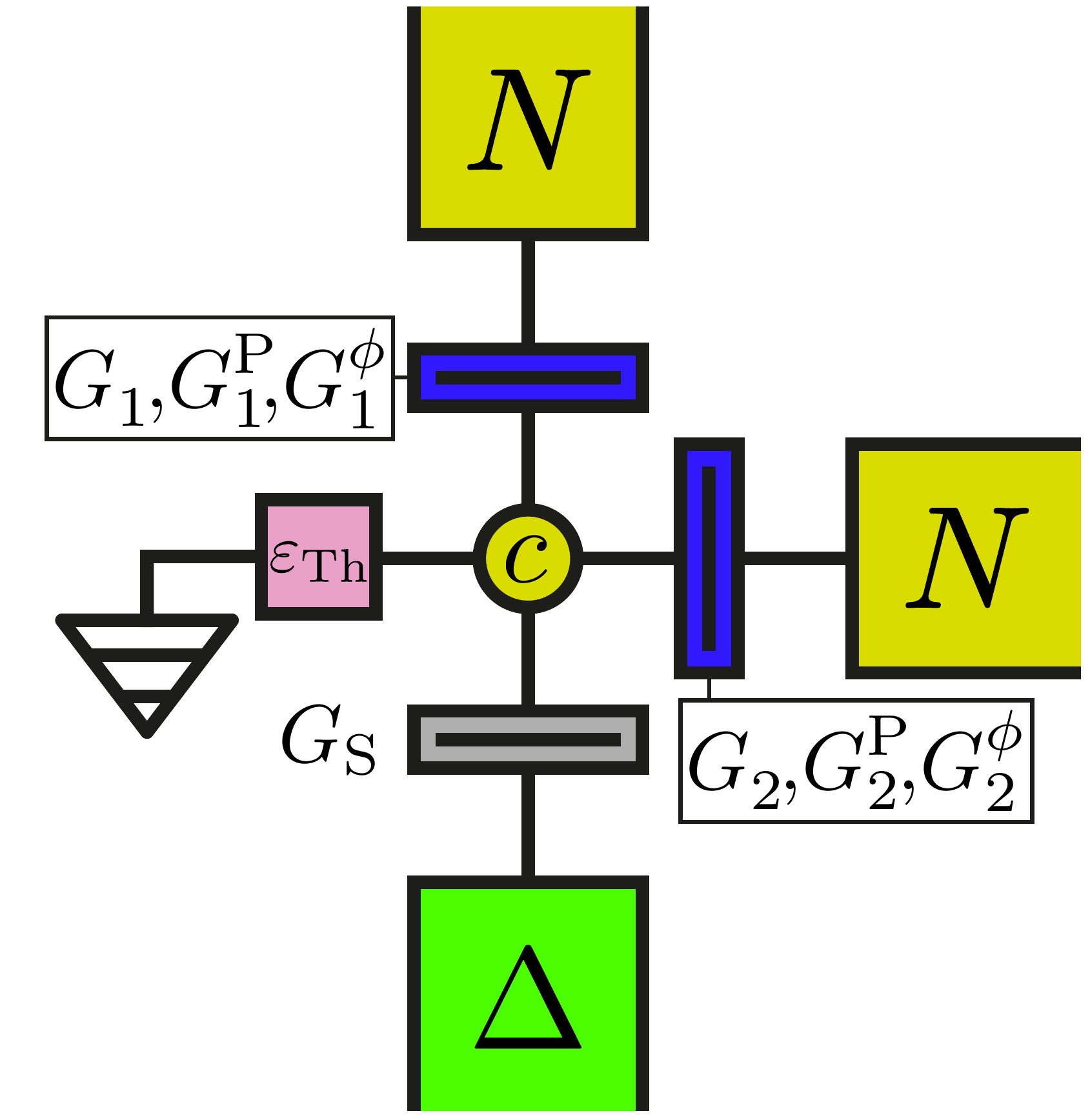}
\put(5,95){\makebox(0,3){$(c)$}}
\end{overpic}
\end{minipage} 
\caption{\label{model}
$(a)$ and $(b)$ show sketches of possible devices to measure the mesoscopic thermopower we predict. The yellow layers are normal metals (N) and the green substrate is the superconductor with gap $\Delta $. 
The spin-dependent effects are induced by a thin ferromagnetic insulator (FI) film shown in blue. The contact between the superconductor and the central normal layer (c) is shown in gray and assumed to be spin inactive. 
To measure the two-terminal effects, the simple layered structure (a) suffices and only the upper normal layer and the superconducting substrate have to be contacted. In (b) an effective two-terminal setup is suggested, where the two normal layers have to be contacted and the superconductor only induces the superconducting correlations but does not need to carry electric current. $(c)$ shows the circuit diagram used to model the setups (a) and (b). The colors are chosen in correspondence to $(a)$ and (b). The normal layer between the superconductor and the ferromagnetic insulating layers is denoted by $c$ and will be called ``node'' in the main text.
The normal N-terminals with FI interfaces could be replaced by ferromagnetic terminals as well, since we choose all energy scales to be small enough, that solely interface effects contribute to the resulting currents. The two-terminal situation is achieved for e.~g. $G_{\mathrm{2}}=G^{\phi}_{\mathrm{2}}=G^{\rm P}_{\mathrm{2}}=0$.}
\end{figure}

The circuit diagram and a sketch of the experimental realization we propose  is depicted in \Fref{model}. The two-terminal case consists of a node [the yellow circle in the center of \Fref{model}(c) denoted `c', and the yellow layer in contact with the superconductor in each of the figures 1(a) and 1(b)] tunnel-coupled to a superconductor (boundary conductance $G_{\mathrm{S}}$) and to a normal metal via a spin-polarized interface (boundary conductance $G_{\mathrm{1}}$), 
e.~g. a thin layer of a ferromagnetic insulator, FI, or, alternatively, via a tunnel contact with a ferromagnetic metal. The role of the node is to harbor the non-equilibrium distribution of quasiparticles as well as the proximity induced superconducting correlations. Its distribution functions and spectral functions will be determined self-consistently by the procedure described below. It is also the place where the spin-polarization of the pair amplitudes takes place via the spin-dependent reflection phases at the ferromagnetic interfaces.

In the effective two-terminal case an additional normal (or ferromagnetic) conductor is coupled to the node ($G_{\mathrm{2}}$), and we assume here that the two contacts are identical ($G_{\mathrm{1}}=G_{\mathrm{2}}$). We call this the ``effective two-terminal case'' since only the two normal leads are contacted. Hence, the setup is similar to a transistor with the normal leads being the source and the drain electrodes. The basis is formed by the contact (via the node) with a proximity induced pair potential through the superconductor. The leakage of coherence in the node is described by an inherent dwell time related to the inverse of the contact's Thouless energy $\varepsilon_{\rm Th}$.

The spin-polarized contacts are quantified by a conductance parameter $G^{\mathrm{P}}=(1/2)(G_\uparrow-G_\downarrow)$, where $G_{\uparrow(\downarrow)}$ is the conductance for spin-up (spin-down) particles (the quantization axis in the contact is given by its interface magnetic moment),
and $G=G_\uparrow+G_\downarrow$. 
The spin-polarized interfaces are described by complex transmission and reflection amplitudes, which are spin-dependent. The necessarily leads to the appearance of spin-dependent scattering phases, also called spin-mixing angles \cite{tokuyasu:88}. The importance of these spin-mixing phases for singlet-triplet conversion in superconductor-ferromagnet heterostructures gives rise to a mechanism for transport through ferromagnets with strong exchange splitting of their bands \cite{eschrig:03}. The necessity to include these effects into quasiclassical theory lead to recent formulations of new sets of boundary conditions for the Eilenberger equation \cite{eschrig:09} and the Usadel equation \cite{cottet:09,machon:13}. An alternative way to arrive at analogous boundary conditions is to introduce a ferromagnetic layer close to the boundary \cite{bergeret:12}.
The spin-dependent phase shifts are the crucial input for obtaining the new type of thermoelectric effects discussed in Ref. \cite{machon:13}.

The spin-dependent scattering phases at the interface give rise to an induced exchange field in the node which is proximity coupled to the superconductor. This effect is quantified by a conductance parameter $G^{\phi_c}$, arising from spin-dependent phase shifts in the reflection amplitudes at the ferromagnetic contacts of the node.
For a relation of the conductance parameters to the microscopic interface properties, see \cite{machon:13}. Note that the description in terms of the quantum circuit theory \cite{nazarov:99,nazarov:book} is very general as only the topology of the circuit enters. E.~g. the magnetic exchange interface can be on the superconducting side in the form of a thin magnetic insulator, a homogeneous (but weak) exchange field in the node or in the terminal, or it can be an externally applied field which does not exceed the critical field of the superconductor (the latter statement was explicitly confirmed in a recent publication \cite{ozaeta:14}). Furthermore, it is easily possible to study more general contacts using generalized boundary conditions for Usadel equations \cite{machon:13}, or to exploit similar effects for the case of ballistic hybrid structures \cite{machon:13,ballistic}.
Hence, the suggested experimental setup is only one possible realization.  

\section{Spin-dependent quasiclassical theory in the dirty limit}
Here we will give a more detailed description of our calculations. In the stationary case the non-equilibrium Keldysh Green function in Fourier presentation $\check{\cal G}(\rv,\rv',\varepsilon)=\int\,dt/\hbar\,\check{\cal G}(\rv,\rv',t-t')\exp{(i\varepsilon(t-t')/\hbar)}$ reads:
\begin{equation}
	\check{\cal G}(\vec{r},\vec{r}^{\,\prime},\varepsilon)=
	\left(\begin{array}[c]{cc}
	\check{\cal G}^{\rm R}(\vec{r},\vec{r}^{\,\prime},\varepsilon) & 
	\check{\cal G}^{\rm K}(\vec{r},\vec{r}^{\,\prime},\varepsilon)\\
	0 & \check{\cal G}^{\rm A}(\vec{r},\vec{r}^{\,\prime},\varepsilon)
	\end{array}\right)
\end{equation}
with
\begin{eqnarray}
	\check{\cal G}^{\rm R/A}(\vec{r},\vec{r}^{\,\prime},t-t^{\prime}) & = &
	\mp i\theta(\pm(t-t^{\prime}))\left\langle \left\{  
	\hat{\Psi}(t,\vec{r}),\hat{\Psi}^{\dagger}(t^{\prime},\vec{r}^{\,\prime})\right\}
	\right\rangle\\ 
	\check{\cal G}^{\rm K}(\vec{r},\vec{r}^{\,\prime},t-t^{\prime}) & = & 
	-i\left\langle \left[  \hat{\Psi}(t,\vec{r}),
	\hat{\Psi}^{\dagger}(t^{\prime},\vec{r}^{\,\prime})\right]\right\rangle.
\end{eqnarray}
Here ${\rm R/A/K}$ labels the retarded, advanced, and Keldysh part respectively.
We choose the spinor basis
\begin{eqnarray}
	\hat{\Psi}\dg(t,\rv)=\left(\Psi\dg\ua(t,\rv),\Psi\dg\da(t,\rv),\Psi\da(t,\rv),-\Psi\ua(t,\rv)\right).
\end{eqnarray}
In the following, $\vec{\check\sigma}$ and $\check\sigma_0$ denote the vector of Pauli matrices and the unit matrix in spin space, respectively, and $\check\tau_z$ denotes the third Pauli matrix in Nambu space. 
In this basis the quasiclassical isotropic Green functions \cite{rammer:86} of a bulk BCS superconductor (S), and a ferromagnet/normal metal (F/N) are given by ($\delta>0$, $\delta\rightarrow0$)
\begin{eqnarray}
	\G_{\mathrm{S}}^{\rm R/A}(\epsilon) & = & 
	\frac{\pm\sign(\varepsilon)}{\sqrt{(\varepsilon\pm i\delta)-|\Delta|^2}}
	\left(\begin{array}{cc}
		(\varepsilon\pm i\delta)&\Delta\\
		-\Delta^*&-(\varepsilon\pm i\delta)
	\end{array}\right)\otimes\check\sigma_0\\
	\G_{\mathrm{N}}^{\rm R/A}(\epsilon) & = & 
	\G_{\mathrm{F}}^{\rm R/A}(\epsilon)=\pm\check\tau_z\otimes\check\sigma_0.\label{gra}
\end{eqnarray}
Here, $\Delta$ is the superconducting order parameter.
Note that the retarded and advanced quasiclassical Green functions for ferro\-magnets and normal metals are equal. The exchange splitting enters in the former case via spin-dependent density of states and diffusion constants, and it enters the boundary condition via spin-dependent phase shifts and interface polarization effects.
Following, e.~g. \cite{belzig:99}, the Keldysh component can be written as
\begin{equation}\label{gk}
	\G^{\rm K}(\epsilon)=
	\G^{\rm R}(\epsilon)\check{h}(\epsilon)-\check{h}(\epsilon)\G^{\rm A}(\epsilon),
\end{equation}
with the distribution matrix (the electrochemical potential of each terminal is measured from the electrochemical potential of the superconductor, with the difference defining the voltage $V$)
\begin{equation}
	\check{h}(\epsilon)=
	\left(\begin{array}{cc}
		\tanh{\frac{\varepsilon-eV}{2T}}&0\\
		0&\tanh{\frac{\varepsilon+eV}{2T}}
	\end{array}\right)\otimes\check\sigma_0.
\end{equation}
Here we have set $k_B=\hbar=1$ and the electric charge to $e=-|e|$. 

We describe the systems shown in \Fref{model} in the discretized version of a quantum circuit theory \cite{nazarov:94,nazarov:99}. The system is divided into terminals, nodes, and connectors. Within a node the Green function $\G_c$ obeys the normalization condition $\G_c^2=1$ and is determined by the boundary conditions in the form of a matrix current conservation, where the form of the matrix current depends on the type of connector.
We use the boundary conditions derived in \cite{machon:13}, which are valid for arbitrary spin polarizations ${\cal P}_n$ for each channel $n$ with the average transmission probability ${\cal T}_n$. The boundary condition depends on the magnetization direction unit vector $\vec{m}$ of the interface, the ferromagnet, or of an external field. It enters the equations below via the matrix $\check{\kappa}=\check\tau_z\otimes(\vec{m}\vec{\check\sigma})$.
In linear order in ${\cal T}_n$ the matrix current between terminals and/or nodes, labeled $j$ and $k$, on the side $k$ of the contact is written explicitly as
\begin{equation}
	\check{\cal I}_{j,k}(\varepsilon)=\frac12
	\left[G^0\check{\cal G}_{j}(\varepsilon)+
	G^{\mathrm{P}}\left\{\check{\kappa},\check{\cal G}_{j}(\varepsilon)\right\}
	+{G}^{1}\check{\kappa}\check{\cal G}_{j}(\varepsilon)\check{\kappa}	
	-iG^{\phi_k} \check{\kappa},\check{\cal G}_{k}(\varepsilon)\right]\,,
\end{equation}
with $\check {\cal G}_j(\varepsilon)$ denoting the Green functions on either side of the connector.
We introduced the conductance parameters
\begin{eqnarray}
	G^0  & = & \frac{e^2}{h} \sum_{n}{\cal T}_n\left(1+\sqrt{1-{\cal P}_n^2}\right)\,,\label{GT} \\
	G^1  & = & \frac{e^2}{h} \sum_{n}{\cal T}_n\left(1-\sqrt{1-{\cal P}_n^2}\right)\,,\label{GMR2}\\
	G^{\mathrm{P}} & = & \frac{e^2}{h} \sum_n{\cal T}_n{\cal P}_n\label{GMR}\,, \\
	G^{\phi_k} & = & 2\frac{e^2}{h} \sum_n \delta \phi_n^k\label{Gfi}\,.
\end{eqnarray}
Here the spin mixing angles $\delta \phi_n^k$ originate from reflection amplitudes on the $k$-side of the interface.
For the setup in \Fref{model}(c) the indices $j$ and $k$ are replaced by a terminal $j\in\{S, 1, 2\}$ and the node $c$, respectively. 

For the connector between the superconductor and the node, $G^1=G^P=G^{\phi_k}=0$ and $G^0=2(e^2/h) \sum_n {\cal T}_n$ (the sum is over the number of channels between the superconductor and the node). Thus, the corresponding boundary condition reads
\begin{eqnarray}
\eqalign{
\check{\cal I}_{S, c}(\varepsilon)=&\frac{G_S}{2}\left[\check{\cal G}_{S}(\varepsilon) ,\check{\cal G}_{c}(\varepsilon)\right]
}
\end{eqnarray}
with $G_S=G_S^0$ is just the standard conductance of the contact to the superconductor.
Here, and in the following we assign to the boundary conductance parameters for the connector between terminal $j$ and the node $c$ for simplicity the index $j$ (remembering that it is always the interface with the node that is considered).

For the ferromagnetic contacts
we have $\left[\G_{j},\hat\kappa_{j}\right]=0$. Thus the terms with $G^0$ and ${G}^{1}$ can be combined into one term with the usual conductance $G=G^0+{G}^{1}=2(e^2/h)\sum\nolimits_{n}{\cal T}_{n}$. The simplified boundary condition reads for $j=1,2$:
\begin{eqnarray}
\eqalign{
\check{\cal I}_{j, c}(\varepsilon)=&\frac{1}{2}\left[G_j\check{\cal G}_{j}(\varepsilon)
+ 2\,G_j^{\mathrm{P}}\check{\kappa}_{j}\check{\cal G}_{j}
(\varepsilon)-iG^{\phi_c}_{j}\check{\kappa}_{j},\check{\cal G}_{c}(\varepsilon)\right].
}
\end{eqnarray}
In the following we will drop for notational simplicity the subscript `$c$' in $\phi_c$, remembering that this spin-mixing parameter refers to the node side of the connector $(j,c)$.
Interestingly, the $G^{\phi}_j$ term can be large compared to $G^0_j$ and $G^1_j$ since channels with ${\cal T}_n=0$ contribute due to spin-dependent reflection phases. On the other hand the $G^{\mathrm{P}}_j$ term is limited, since $G^{\mathrm{P}}_j=(1/2) (G_{j\uparrow}-G_{j\downarrow})=P_jG_j/2$ with the polarization $P_j=(G_{j\uparrow}-G_{j\downarrow})/(G_{j\uparrow}+G_{j\downarrow})$, and thus $P_j\in[-1,1]$.

The loss of superconducting correlations in the node region is described by a virtual leakage terminal,
\begin{eqnarray}
\eqalign{
\check{\cal I}_{\rm Leak}(\varepsilon)=-\frac{e^2}{h}\frac{i\varepsilon}{2\varepsilon_{\rm Th} }
[\check\tau_z\otimes\check\sigma_0,\G_{c}(\varepsilon)],
}
\end{eqnarray}
with
the Thouless energy $\varepsilon_{\rm Th}$. Note that the Thouless energy in thin layers as we discuss here depends on the contact conductances and we will therefore adapt the Thouless energy to match the appropriate mini gap in the density of states.

The Green function of the contact region, $\G_c$, is determined by the matrix current conservation  \cite{nazarov:book}
\begin{eqnarray}
\eqalign{
\check{\cal I}_{\rm Leak}(\varepsilon)+\sum_{j\in (S,1,2)}\check{\cal I}_{j,c} (\varepsilon) =0, 
}
\end{eqnarray}
which can be written in the form $[\check{\cal M},\check{\cal G}_c]=0$. 
We solve this equation by diagonalizing $\check {\cal M}$ via $\check {\cal M}=\check U $diag$(\check {\cal M}) \check U^{-1}$ and then calculate $\check {\cal G}_c=\check U $sign$[$Re~diag$(\check {\cal M})] \check U^{-1}$.
This ensures that $\check{\cal G}_c^2=\check 1$, and that the eigenvalues $\pm 1$ of $\check{\cal G}_c$ are continuously connected to those of the normal state solution (Equations \ref{gra} and \ref{gk}) 

Knowing $\check{\cal G}_c$ we calculate all matrix currents $\check{\cal I}_{j,c}$. The charge ($I^q$) and energy currents ($I^{\varepsilon}$) are now obtained from the Keldysh component of the matrix current \cite{morten:06}:
\begin{eqnarray}\label{current:dirty}
I^q_{j}&=\frac{1}{8e}\int\mathrm{Tr}\left[
(\check{\tau}_z\otimes\check{\sigma}_0)\,
\check{\cal I}_{j,c}^{\scriptstyle{\rm K}}(\varepsilon) 
\right] d\varepsilon\,,\\
I^{\epsilon}_{j}&=\frac{1}{8e^2}\int \varepsilon\,\mathrm{Tr}\left[\check{\cal I}_{j,{\rm  c}}^{\scriptstyle{\rm K}}(\varepsilon)\right]d\varepsilon\,.
\end{eqnarray}
The density of states in the contact region is obtained from the retarded Green function ${\cal G}^{\rm R}_c$ like
\begin{eqnarray}
N(\varepsilon)=\frac{1}{4}\Re{\,\mathrm{Tr}[(\check\tau_z\otimes\check\sigma_0)\,\check{\cal G}^{\rm R}_c(\varepsilon) ]} .
\end{eqnarray}

\section{Thermopower}\label{TP}
To keep things simple, in the following we will consider the linear response coefficients only. With Einstein summation convention the corresponding currents are ($i,j \in [1,n]$, with the number of normal terminals $n=1,2$)
\begin{equation}\label{lmat}
\label{condmat}
	\left(\begin{array}{c} I^q_i\\I^{\varepsilon}_i\end{array}\right)=
	\left(\begin{array}{cc} L^{qV}_{ij}&L^{qT}_{ij}\\
	L^{\varepsilon V}_{ij}& L^{\varepsilon T}_{ij}\end{array}\right)
	\left(\begin{array}{c}e V_j \\ \Delta T_j/T\end{array}\right)\,.
\end{equation}
Here $\Delta T_j=T_j-T$ and $e V_i$ are the temperature and the electrochemical potential differences measured relative to the temperature $T$ and the electrochemical potential of the superconductor, respectively. Note that the Onsager relation \cite{onsager} is satisfied since the conduction matrix is fully symmetric \cite{machon:13}. 

\begin{figure}
\centering
(a)\includegraphics[height=0.45\linewidth]{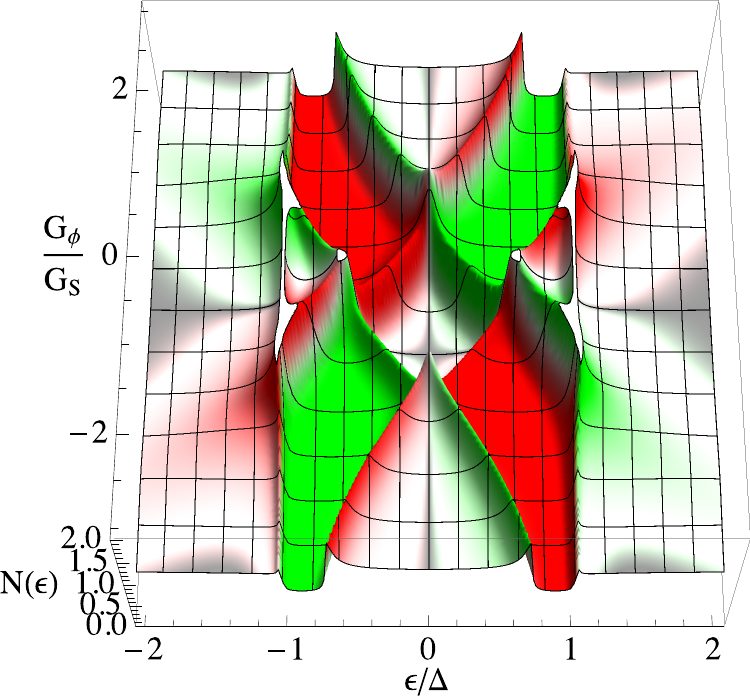}
(b)\includegraphics[height=0.45\linewidth]{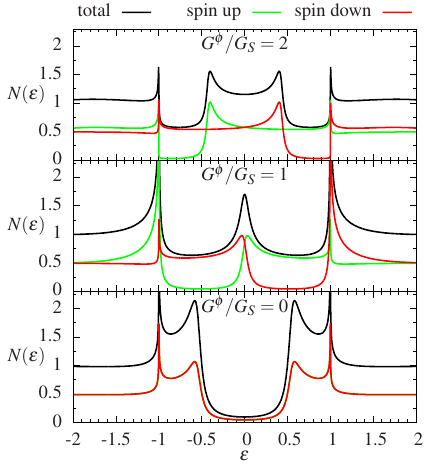}
\caption{\label{dos} 
Density of states (DOS) $N(\varepsilon)$ at $T=0.5\,T_c$ in the node of the two-terminal contact for
$G_{\rm 1}=0.1 G_{\rm S}$ and the Thouless energy $\varepsilon_{\rm Th}=\Delta_0$, with $\Delta_0=\Delta(T=0)$.
In (a) a surface plot of
the total DOS depending on $G^{\phi}$ is shown. The color code describes the spin polarization (green/red for 100\% spin up/down). In (b) the DOS and the SDOS are shown for three selected values of $G^{\phi}$ (all curves are normalised to the normal state DOS).
}
\end{figure}

The thermoelectric effects in our setups can be qualitatively understood from the spin-dependent density of states (SDOS). In \Fref{dos}(a) we plot the SDOS in the node, for the two-terminal case. The plot shows the total density of states with colors encoding the spin polarization from red (100\% spin down) to green (100\% spin up). Without spin active interface ($G^{\phi}=0$) we observe a peak at the superconducting gap $\Delta=\Delta(T)$ and another, broader peak defining the so-called minigap, which is of the order of the Thouless energy. This is also shown in the lowest panel of \Fref{dos}(b). Here all curves are normalized to the normal state DOS. The effect of the spin-active interface parameter $G^{\phi}$ is to spin-split the subgap states similarly as in the presence of a Zeeman interaction or an exchange field. 
As can be seen in \Fref{dos}(a) for increasing $G^{\phi}$ the spin-dependent minigaps shift in opposite directions in energy for the two spin species. Thus, with increasing spin mixing the minigap in the total DOS shrinks, resulting in two bands of almost fully spin-polarized states (the green and red regions in \Fref{dos}(a)). These plateaus are formed by the continuum of one spin species inside the minigap of the other one. As we can see they are bound by one (and above the a critical value $G^{\phi}$ where the minigap closes by two) peak(s) with opposite spin-polarization. At $G^{\phi}=\Delta/\epsilon_{\rm Th}$ the peaks beyond the plateau cross the $\Delta$ peak as can be seen also in the middle panel of \Fref{dos}(b). For higher values of $G^{\phi}$ the spin-polarized bands shrink as the minigap peaks within the superconducting gap asymptotically approach the peak at $\Delta$. A similar picture of the DOS and of the SDOS for the case with three terminals has been discussed in \cite{machon:13}. Thus, $G^{\phi}\neq 0$ leaves the total DOS symmetric but breaks electron-hole symmetry for each spin direction. As we show below combining this with spin-dependent tunneling (i.e $G^{\rm P}\neq 0$) results in a finite thermoelectric effect, i.e. a nonzero $L^{qT}$.

In the case of a pure two-terminal device, \Eref{condmat} describes the total current through the junction. Therefore we define the thermopower in this case as $S=-V_1/\Delta T_1=(1/eT) \,L^{qT}_{11}/L^{qV}_{11}$ using the usual open circuit boundary condition $I^q=0$. In the effective two-terminal case we assume that only the two normal leads (see \Fref{model}) are contacted and the superconductor is used to induce proximity amplitudes in the normal conducting node. This already results in the condition $I^q_1=-I^q_2$ for the charge current. Motivated by recent experiments \cite{giazotto:12} we will assume a temperature difference ($\Delta T_i$ where $i\in\{1,2\}$ labels the side) on one side of the junction. The resulting thermopower over the whole structure can be divided in two contribution, defined as ($V=V_1-V_2$):
\begin{eqnarray}\label{eff}
S_{i}=(-1)^{i}\frac{V}{\Delta T_i}\Bigg|_{I^q_1=0}=
(-1)^{i}\frac{1}{T}\frac{L_{\rm 2i}^{\rm qT} (L_{\rm 11}^{\rm qV} + L_{\rm 12}^{\rm qV}) - L_{\rm 1i}^{\rm qT} (L_{\rm 12}^{\rm qV} + L_{\rm 22}^{\rm qV}) }{L^{\rm qV}_{\rm 11} L^{\rm qV}_{\rm 22}-(L^{\rm qV}_{\rm 12})^2}.
\end{eqnarray}
Here we have again used the open circuit boundary conditions $I^q_1=I^q_2=0$. The two contributions $S_1$ and $S_2$ are generally not equal since they depend on the magnetization strength and directions as well as on the tunneling conductance of both interfaces. We assume here for simplicity that both contacts to terminal 1 and 2 are characterized by the same parameters ($G^{\rm P}_1=G^{\rm P}_2=G^{\rm P}$, $G^{\rm \phi}_1=G^{\rm \phi}_2=G^{\rm \phi}$, $\check{\kappa}_1=\check{\kappa}_2$, and $G_{\rm 1}=G_{\rm 2}$). For this choice the Onsager matrix in \Eref{lmat} has more symmetries that dictated by the Onsager symmetry alone (e.~g. $L^{\rm qT}_{21}=L^{\rm qT}_{12}$ and $L^{\rm qT}_{11}=L^{\rm qT}_{22}$). As consequence we obtain $S=S_1=S_2$.

\begin{figure}[t]
\centering
(a)
\includegraphics[height=5.5cm]{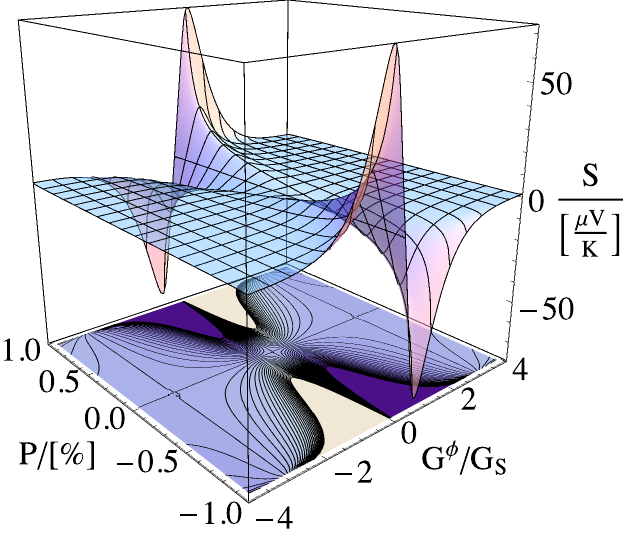}
(b)
\includegraphics[height=5.5cm]{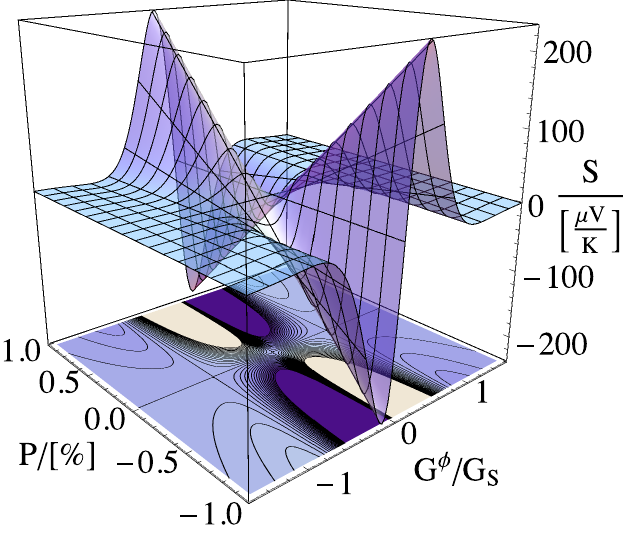}
\caption{
	Thermopower for the two-terminal $(a)$ and the effective two-terminal $(b)$ case, depending on the interface spin-mixing parameter $G^{\phi}$ (providing an effective Zeeman splitting) and on the interface polarization $G^{\rm P}$. In ($a$) we choose $G_1=0.1\,G_{\rm S}$ and in ($b$) both contacts are chosen to be equal at the ratio $G_{\rm 1/2}=0.01\,G_{\rm S}$. The other parameters are $\varepsilon_{\mathrm{Th}}=2\Delta_0$ and $T=0.5\,T_c$ in panel ($a$) and  $\varepsilon_{\mathrm{Th}}=\Delta_0$ and  $T=0.2\,T_c$ in panel ($b$).}
	\label{s_gmr_gphi} 
\end{figure}

\begin{figure}[t]
	(a)\includegraphics[height=5.5cm]{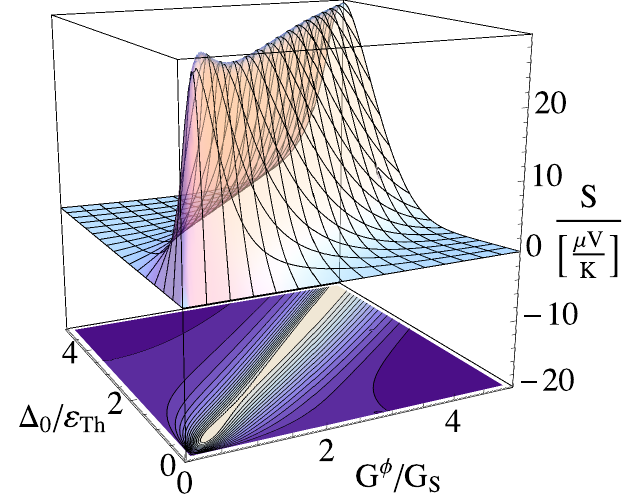}
	(b)\includegraphics[height=5.5cm,unit=1mm]{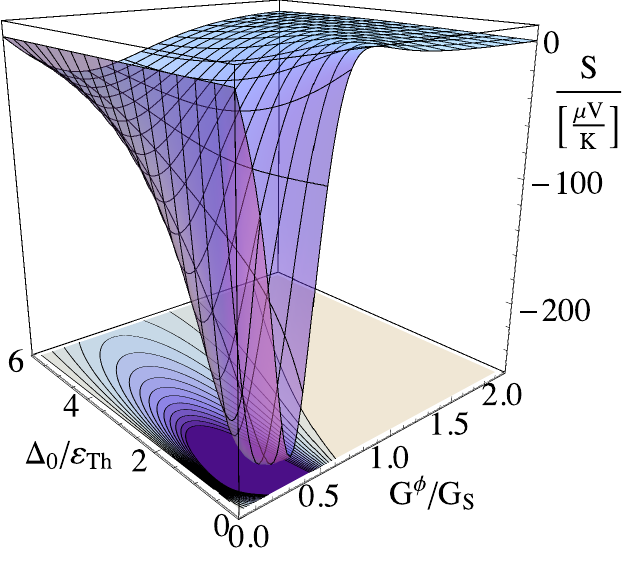}
	\caption{\label{s_eth_gphi} 
	Thermopower  depending on the Thouless energy $\varepsilon_{\rm Th}$ and on the spin-mixing parameter $G^{\phi}$ (providing an effective Zeeman splitting) for ($a$) the two-terminal and ($b$) the effective two-terminal case. In both cases $P=0.9$, the temperature and the conductance are in ($a$) $T=0.5\,T_c$, $G_1=0.1 G_{\rm S}$ and in ($b$) $T=0.2\,T_c$, $G_{1/2}=0.01 G_{\rm S}$.}
\end{figure}

In \Fref{s_gmr_gphi} we show the dependence of the thermopower on the polarization of the interface $G^{\mathrm{P}}$ and the spin-mixing effect $G^{\phi}$ for both considered setups. 
We have chosen slightly different parameters in both plots to obtain the maximal Seebeck coefficient in each case. 
For the two-terminal case (\Fref{s_gmr_gphi}a) we observe a strong increase of the thermopower for large polarizations, while the thermopower in the effective two-terminal (\Fref{s_gmr_gphi}b) case depends linear on $G^{\mathrm{P}}$.
This is expected since in the two-terminal case only quasiparticles above the superconducting gap can contribute to an energy current into the superconductor. Hence, for low temperatures the nonzero coefficients $L^{qT}=L^{\varepsilon V}$ in the two terminal case are solely due to the  small fraction of excitations above the superconducting gap. On the other hand $L^{qV}$ is mainly the Andreev conductance, which is predominantly due to transport at energies below the superconducting gap. For a large interface polarization Andreev processes are strongly suppressed, 
leading to an increase of the thermopower. The maximal Seebeck coefficient is reached roughly at $G^{\phi}/G_S=\Delta/\varepsilon_{\rm Th}$ as can be seen even better in \Fref{s_eth_gphi}a. This is understood from \Fref{dos}, since this corresponds to the point when the outer subgap peak (dispersing from the peaks defining the minigap at $G^{\phi}=0$ towards the superconducting gap edges) in the DOS crosses the peak at the superconducting gap $\Delta$ and hence contributes to thermal transport into the superconductor. 
In the limit of Thouless energies $\epsilon_{\rm Th}\gg 1$ the unpolarised minigap at $G^{\phi}=0$ merges with the superconducting gap and thus also results in a vanishing Seebeck coefficient.

The Seebeck coefficients in \Fref{s_gmr_gphi}b and \Fref{s_eth_gphi}b are much larger over a wider range of parameters.
The reason is another mechanism that creates the thermoelectric response.
In the pure two-terminal case only quasiparticles above the superconducting gap contributed to the thermopower. In the effective two-terminal case, when both leads are normal, also subgap states can contribute to the thermal transport. 
The main contribution comes from the strong energy asymmetry in the almost fully spin-polarized bands (red and green regions in \Fref{dos}). Hence, the thermopower is carried by a rather broad spin-polarized energy band, which is driven by the quasiparticles in the most important energy range $~k_BT$. This is one of our main results and justifies to study the more complicated lateral structure which has to be used for the effective two-terminal geometry. 

Now the different signs of the dominating peak in the Seebeck coefficient between the two-terminal and the effective two-terminal cases becomes clear. If one e.~g. considers only the upper half-plane (positive $G^{\phi}$) in \Fref{dos}, the main asymmetry in spin is created from the plateaus which are formed by spin-up particles (green) for positive energies and spin-down particles (red) for negative energies. But in the pure two terminal case only quasiparticles contribute with $|\epsilon| > |\Delta|$. Their spin is the opposite (down for positive and up for negative energies) compared to the once forming the plateau. Hence, for equal polarization the thermoelectric coefficients have a different sign in the two cases.
In the same way the sign of the second small peak in the $G^{\phi}$ dependence of \Fref{s_gmr_gphi}b is understood: In the DOS in \Fref{dos} at $\epsilon=0.2\Delta$ (which roughly matches the relevant energy range at $0.2\,T_c$) one sees for increasing $G^{\phi}$ that  the DOS first is unpolarized, reaches a rather broad spin-up  plateau, then a small spin-down peak, and finally for $\epsilon\gg\Delta$  both spin directions become equal again. This exactly matches the first large peak and a second very small peak with opposite sign. 
In \Fref{s_eth_gphi}b the thermopower for a fixed polarization is shown. The maximum is found roughly at $\epsilon_{\rm Th}=0.5\Delta$ and $G^{\phi}=0.25 G_{\rm S}$. Here the plateau in the SDOS becomes maximal. We considered $T=0.2\,T_c$ since higher temperatures lead to an increasing contribution of the quasiparticles above the minigap which produce a thermoelectric effect with opposite sign with respect to the contribution from the plateau, and hence reduce the total effect. 
Note that only in the two-terminal case the thermopower vanishes for $\varepsilon_{\rm Th}\rightarrow\infty$, whereas the spin-polarized plateaus in the effective two-terminal case remain. This is in agreement with the statement of a vanishing local thermopower for two terminals in the clean limit \cite{machon:13}.

\section{Figure of Merit}
The commonly used definition of the figure of merit is $ZT=GS^2T/\kappa=GS^2T/(\kappa_0-GS^2T)$ with the conductance $G$ and the zero current thermal conductance $\kappa$. In the two terminal case we have $G=eL^{qV}_{11}$, $\kappa_0=L^{\varepsilon T}_{11}$ and the definition of $S$ given in the previous section. Then we can express the figure of merit through the linear response coefficients as
\begin{eqnarray}
ZT=\frac{(L^{qT}_{\rm 11})^2}{L^{qV}_{\rm 11} L^{\varepsilon T}_{\rm 11}-(L^{qT}_{\rm 11})^2}.
\end{eqnarray}
Due to the second law of thermodynamics, the conduction matrix in \Eref{condmat} is positive definite \cite{onsager,bergmann:91}. Accordingly, the figure of merit takes values between zero and infinity where infinity corresponds to Carnot efficiency. 

In the effective two-terminal system we use the same conditions as in the previous section. From the condition that the superconductor is electrically isolated we obtain the conductance for the whole junction as
\begin{equation}
	G = \frac{I^{\rm q}_{\rm 1}}{V} 
	=\frac{L^{\rm qV}_{\rm 11} L^{\rm qV}_{\rm 22}-(L^{\rm qV}_{\rm 12})^2}{
		L^{\rm qV}_{11}+2L^{\rm qV}_{12}+L^{\rm qV}_{22}}.
\end {equation}
Now we assume a temperature difference $\Delta T_1\neq0$ and hence use the definition $S_1$ for the thermopower. Since the leakage of heat current to the superconductor and the substrate cannot be controlled easily (even though the superconductor is a good thermal insulator at low temperatures) we would like to avoid definition requiring a control of thermal currents. Therefore the choice of the thermal conductance is not unique. 
The energy currents at zero charge current in the normal terminals can be expressed in terms of the previously defined thermopowers like
\begin{eqnarray}
I^{\rm \epsilon}_i=\kappa_i\Delta T_1=(-1)^i(G S_i S_1 T-\kappa_{0i})\Delta T_1.
\end {eqnarray}
Here we defined 
\begin{eqnarray}
\kappa_{0i}=\frac{(-1)^i}{T}\left(\frac{(L^{\rm qT}_{1i}+L^{\rm qT}_{2i})(L^{\rm qT}_{11}+L^{\rm qT}_{21})}{L^{\rm qV}_{11}+2L^{\rm qV}_{12}+L^{\rm qV}_{22}}-L^{\rm \epsilon T}_{i1}\right).
\end {eqnarray}
One possible choice is to define $\kappa$ via $I^{\rm \epsilon}=-I^{\rm \epsilon}_2=\kappa\Delta T_1$, which is very much in the sense of the mentioned experiment \cite{giazotto:12}. This results in the figure of merit
\begin{eqnarray}\label{zt1}
ZT=\frac{GS_1^2T}{\kappa_{02}-GS_1S_2T}.
\end {eqnarray}
Another reasonable choice is $I^{\rm \epsilon}=(I^{\rm \epsilon}_1-I^{\rm \epsilon}_2)/2=\kappa\Delta T_1$ which accounts for the energy current which is flowing from terminal 1 to terminal 2. 
In this case the figure of merit is given by
\begin{eqnarray}
ZT=\frac{GS_1^2T}{(\kappa_{01}+\kappa_{02})/2-GS_1T(S_1+S_2)/2}.
\end {eqnarray}
Since we have symmetric contacts, in our case $\kappa_{01}$ and $\kappa_{02}$ differ only in the parameter $L^{\epsilon T}_{i1}$. In the following we choose the first definition (\ref{zt1}) for all plots.

\begin{figure}
\centering
(a)\includegraphics[height=5.5cm]{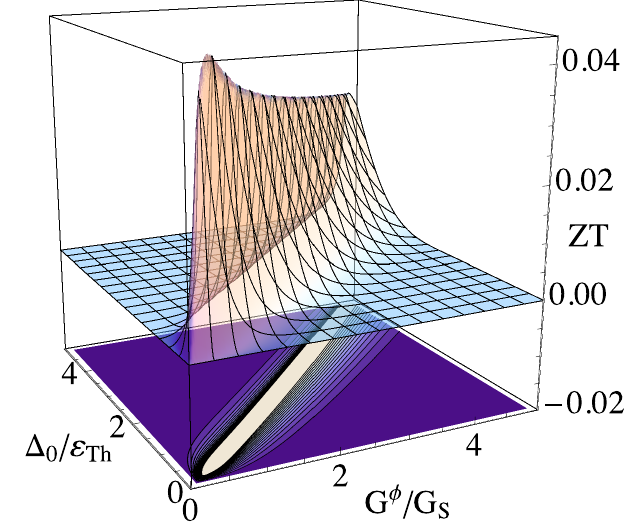}
(b)\includegraphics[height=5.5cm]{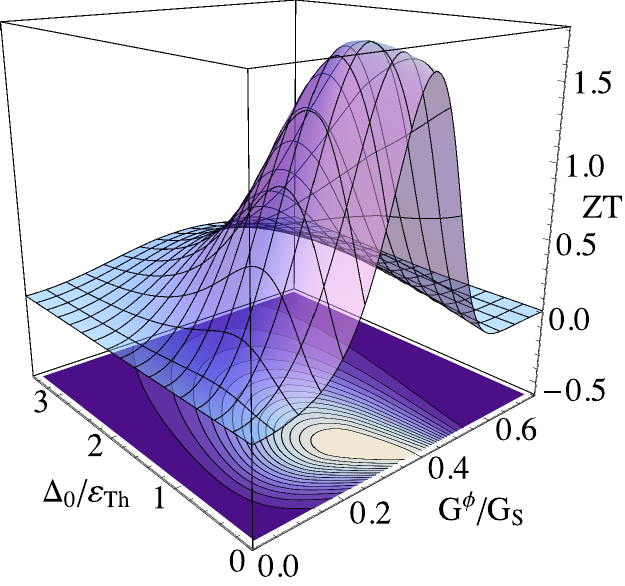}
\caption{\label{zt_eth_gphi} 
Thermoelectric figure of merit for the local transport parameters at the ferromagnetic contact, depending on the strength of the Thouless energy and on the 
spin-mixing parameter $G^{\phi}$ (which causes an effective Zeeman splitting)
for ($a$) the two-terminal and ($b$) the effective two-terminal case. The parameters are chosen as $P=0.9$ and $G_{\rm 1/2}=0.01\,G_{\rm S}$ the temperature is ($a$) $T=0.5\,T_c$ and ($b$) $T=0.2\,T_c$.}
\end{figure}

\Fref{zt_eth_gphi} shows the figure of merit of both setups depending on the strength of the spin-mixing parameter and the Thouless energy. Not surprisingly, the figure of merit reaches it's maximum for roughly the same parameters as the thermopower (compare to \Fref{s_eth_gphi}). 
In the two terminal case 
the maximum appears in \Fref{zt_eth_gphi} for relatively large Thouless energies (i.~e. $\epsilon_{\rm Th} \sim \Delta$). The reason is that the peaks defining the minigap for this case are shifted close to the superconducting gap $\Delta$ and therefore small values of $G^{\phi}$ can already shift a considerable part of the spin-polarized peak to energies larger than the superconducting gap (see \Fref{dos}). 
Consquently for lowering the Thouless energy the peaks defining the minigap are shifted towards zero energy and correspondingly larger spin splitting is required to maximize  $ZT$. Overall, we note that the maximal $ZT$ in our two terminal setup is $\lesssim 0.05$ and therefore comparatively low.

In the view of thermoelectric figure of merit, the effective two-terminal case is much more interesting. Here a peak in $ZT$ appears with it's height reaching and even exceeding $1.5$ (see \Fref{zt_eth_gphi}b). This maximum is achieved at the same parameters for which the thermopower is also maximal. 
An important parameter to optimize $ZT$ is the coupling strength $G_{\mathrm{1/2}}/G_{\mathrm{S}}$, since weaker coupling of the normal metals leads to sharper peaks and more pronounced gaps in the DOS. It is generally known \cite{mahan:96} that the combination of a narrow distribution functions and a high amount of charge and energy carriers in the peaked region of the DOS leads to high efficiencies. Therefore we have chosen rather weakly coupled normal leads. 
We can achieve much higher values of $ZT$ than in the two terminal case since all the highly polarized subgap density of states contributes to the current. The optimal values of Thouless energy and spin mixing are explained by the fact that they on one hand maximize the width of the spin polarized energy bands in the DOS of the node and on the other hand shift them into the optimal energy window $~k_BT$. It is interesting to note that even for very small nodes, i.~e. large Thouless energy a sizable figure of merit remains. In this limit the relevant energy for the spin splitting is given by the superconducting gap $\Delta$.

\section{Temperature Dependence}

\begin{figure}[t]
\centering
\begin{overpic}[width=\linewidth]{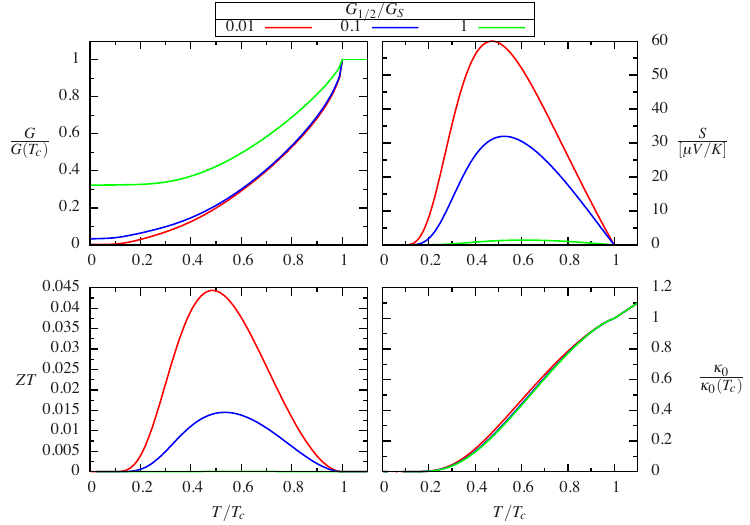}
\put(15,60){\makebox(0,3){$(a)$}}
\put(54,60){\makebox(0,3){$(b)$}}
\put(15,27){\makebox(0,3){$(c)$}}
\put(54,27){\makebox(0,3){$(d)$}}
\end{overpic}
\caption{\label{matrix}
Temperature dependence of the independent linear response coefficients 
and the corresponding thermopower 
in the two-terminal case for $\epsilon_{\rm Th}=2\Delta_0$, $P=0.9$ and $G^{\phi}=0.5\,G_{\rm S}$. 
(a) and (d) show the usual electrical ($G$) and thermal conductances ($\kappa_0$). (b) shows the Seebeck coefficient $S$, and (c) the thermoelectric figure of merite $ZT$.
$G=L^{qV}_{11}$ and $\kappa_0=L^{\epsilon T}_{11}/T$ are normalized to their normal state values $G(T\ge T_c)=\sum_{\sigma=\uparrow,\downarrow}G_{\rm S}G_{\sigma}/(G_{\rm S}+2G_{\sigma})$ and $\kappa_0(T_c)=G(T_c)T_c(\pi^2/3)$. The thermopower is shown in units $\mu V/K$. The color code labels different couplings $G_{\rm 1}/G_{\rm S}$.}
\end{figure}

\begin{figure}[t]
\centering
\begin{overpic}[width=\linewidth]{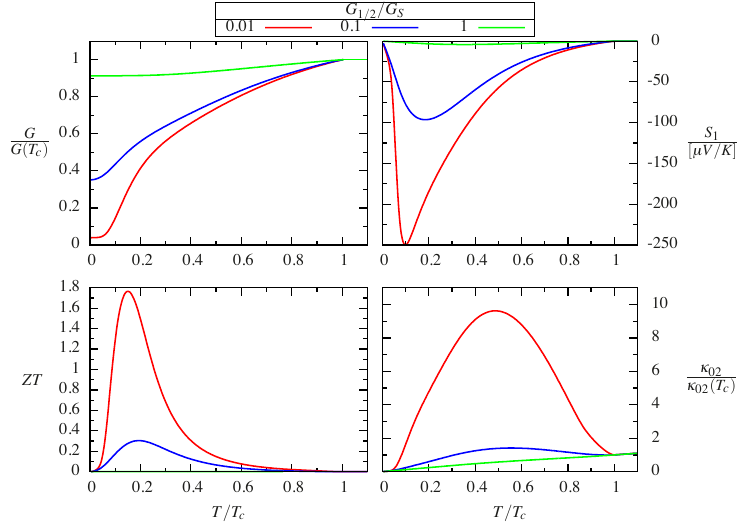}
\put(15,60){\makebox(0,3){$(a)$}}
\put(54,60){\makebox(0,3){$(b)$}}
\put(15,27){\makebox(0,3){$(c)$}}
\put(54,27){\makebox(0,3){$(d)$}}
\end{overpic}
\caption{\label{matrix_eff}
Temperature dependence ($T/T_c$ with $T\equiv T_{\rm S}$) of the effective transport coefficients, the thermopower and the figure of merite in the effective two terminal case 
for $\epsilon_{\rm Th}=\Delta_0$ and $G^{\phi}=0.3\,G_{\rm S}$.
(a) and (d) show the effective electrical and thermal conductances. (b) shows the thermopower, and (c) the figure of merit.
The conductances are normalized to their normal state values. The thermopower is shown in units $\mu V/K$. The color code labels different couplings $G_{1/2}/G_{\rm S}$.}
\end{figure}
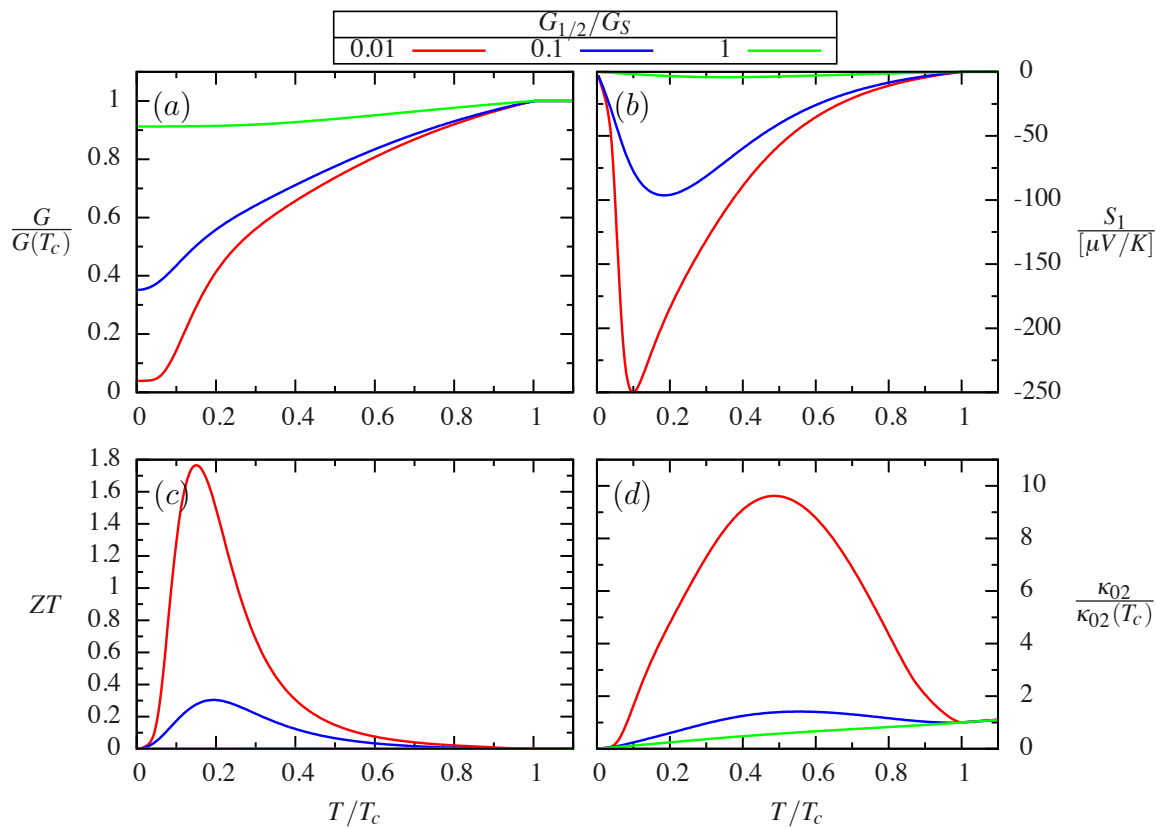

Following the argumentation of \Sref{TP} we expect a strong temperature dependence of the response parameters. In \Fref{matrix} we plot the temperature dependence of the independent conductances ($G,\kappa_0$), the local thermopower, and the figure of merit in the two terminal case. \Fref{matrix} (a) and (d) show the usual electrical and thermal conductances. Both decrease with lowering the temperature, but due to the Andreev contribution the electrical conductance $G=L^{qV}_{11}$ remains finite at very low temperatures. With increasing the coupling $G_1$ to the normal leads the Andreev conductance increases. The thermopower ($S$) and the figure of merit ($ZT$) have a non-monotonous temperature dependence, as shown in \Fref{matrix} (b) and (c), respectively. They obviously vanish in the low temperature limit as well as above the critical temperate $T_c$. To increase the maximal thermopower it is favorable to choose a small coupling $G_1$ to the normal terminal. Physically this can be understood, since a stronger coupling $G_1$ leads to a smearing of the sharp features in the spin-resolved densities of states and, hence, to a reduction of the effective spin-polarization, which is responsible for the thermoelectric effect in our device. The maximum of $L^{\rm qT}_{11}$ (not shown) is roughly at a temperature when $\Delta(T)$ matches the minigap. For our parameter set this is achieved at $T\sim0.7\,T_c$. But the maximum of $S$ is reached below since the conductance is rising for higher temperatures.

In \Fref{matrix_eff} we plot the temperature dependent thermoelectric response coefficients, the effective conductance parameters ($G,\kappa_{02}$) and the thermoelectric figure of merit in the effective two terminal geometry. We observe that the conductance is much less suppressed than in the two terminal case. The reason is that charge transport for subgap energies here also is due to quasiparticles, which can propagate between the two normal terminals. A pronounced thermopower and large figure of merit is achieved for comparatively low temperatures around $0.1-0.2\,T_c$. At the same time, a large peak appears in the heat conductance below $T_c$ for small values of $G_{1/2}/G_{\rm S}$.

As pointed out before, we expect different temperature dependences of the thermoelectric response in the  two setups shown in \Fref{model} (a) and (b), since in the effective two terminal case two contributions are competing. The main one is coming from the almost fully spin-polarized plateaus seen in the DOS (\Fref{dos}, green and red bands), and the second one comes from the minigap peaks and has the opposite sign due to the opposite spin polarization. Since the second contribution becomes stronger for higher temperatures and at the same time the conductance increases, the maximal thermopower is achieved at lower temperatures, compared to the pure two terminal case.
Another factor reducing the figure of merit $ZT$ for temperatures $\sim 0.5\,T_c$ (especially for $G_{1/2}/G_S=0.01$) is the high value of the thermal conductance, that is due to the strongly peaked DOS around energies of the order of $k_BT$. Still, for $T\sim0.15\,T_c$, $ZT$ reaches values $\sim1.8$.   

\section{Conclusions}
We have shown that large thermoelectric effects can be achieved in mesoscopic superconductor-ferromagnet heterojunctions. We have studied two realistic setups. One is a two terminal device where the thermoelectric effects are solely due to quasiparticle excitations above the superconducting gap. 
The second setup is an effective two terminal device with normal leads allowing for subgap energy currents. 
Interestingly, in both cases we found large values of the Seebeck coefficient for strong interface spin polarizations. 
With a view on possible applications we calculated the figure of merit. For the effective two terminal setup we find large 
efficiencies for reasonable interface parameters. In the pure two terminal case such high efficiencies cannot be achieved since only weakly spin-polarized quasiparticles above the superconducting gap contribute. We believe our suggested setups can be readily realized combining ferromagnetic insulator heterostructures \cite{huebler:12} and local caloric techniques \cite{giazotto:12}.

\section*{Acknowledgements}
 
WB and PM acknowledge financial support from the DFG through SFB 767 and BE 3803/03
and from the Baden-W\"urttemberg-Stiftung. ME acknowledges support from the EPSRC under grant reference EP/J010618/1. ME and WB were supported from the Excellence Initiative program ``Freir\"aume f\"ur Kreativit{\"a}t'' at the University of Konstanz.

\section*{References}

\bibliographystyle{phaip}
 
\end{document}